\documentstyle[12pt]{article} 

\def\lp{\stackrel{\leftarrow}{\partial}}
\def\rp{\stackrel{\rightarrow}{\partial}} 
\def\bowstar{\bowtie\kern-0.8em |~}

\begin{document}
\vspace{-2.0cm}
\bigskip

\begin{center}
{\Large \bf
On the Associativity of Star Product in
Systems with Nonlinear Constraints}
\vskip .8 true cm
{\bf Rabin Banerjee}\footnote{rabin@bose.res.in},
{\bf Biswajit Chakraborty}\footnote{biswajit@bose.res.in} and
{\bf Tomy Scaria}\footnote{tomy@bose.res.in} 
\vskip 1.0 true cm

S. N. Bose National Centre for Basic Sciences \\
JD Block, Sector III, Salt Lake City, Calcutta -700 098, India.
\end{center} 
\bigskip
 
\centerline{\large \bf Abstract}
The noncommutative star product of phase space functions is, by construction, 
associative for both non-degenerate and degenerate case (involving only second 
class constraints) as has been shown by Berezin, Batalin and Tyutin. However,
for the latter case, the manifest associativity is lost if an arbitrary 
coordinate system is used but can be restored by using
an unconstrained canonical set. 
 The existence of such a  canonical transformation is 
guaranteed  by a theorem due to Maskawa and Nakajima. 
In terms of these new variables, the  Kontsevich series for the
star product reduces to an exponential series which is manifestly associative. 
We also show, using the star product formalism, that the angular momentum of a particle
moving on a circle is quantized. 
 
\section{Introduction}
Noncommutative geometry\cite{connes} is one of the most important areas of investigation 
these days. It is also related to deformation quantization
\cite{bayen} where ordinary 
commutative multiplication between c-number valued functions of classical
 phase space variables is replaced by noncommutative star product which is
supposed to be associative in nature. For a simple classical system like 
a particle moving on a real line, the construction of the star product can be 
 motivated by considering the set of Weyl ordered phase space operators
and its isomorphism to the set of classical phase space functions. (See \cite{zachos} for a review.) The rule
of composition of the phase space operators is ordinary multiplication of
Weyl ordered operators whereas it is given by the star product in the space of c-number
valued classical functions.This has been shown by Groenewold
\cite{groenewold, moyal} and later through a geometric approach by Berezin \cite{berezin}, Batalin and Tyutin \cite{batalin}. Recent studies
in string theory also indicate that at a very small length scale the nature
of space-time co-coordinates may become noncommutative \cite{witten, szabo, dn, bcg}. 

One  problem that plagues the definition of star product is that it is not 
manifestly associative. This is despite the fact that the star product, by construction, is taken to be associative
both in the non-degenerate and degenerate cases (i.e., in presence of second class constraints) in the geometric 
approach of Batalin and Tyutin \cite{batalin}.
 Since the star product is of fundamental importance
in defining noncommutative space-time, it is essential to make a reappraisal 
of the problem of associativity of this product. Also it becomes important to compare the 
non-geometrical approach of Groenewold 
\cite{zachos, groenewold} with the geometrical approach \cite{batalin}.

In the above mentioned example of a particle moving 
on the real line ${\cal R}^1$ the classical phase space is given by the 2-d
space  ${\cal R}^2$. As has been shown in \cite{zachos},  the star product between 
two phase space functions $f(x, p)$ and $g(x, p)$ can be expressed in a
Fourier representation by  an exponential factor involving the 
area of a triangle
so that the associative nature becomes manifest.
However, this property of associativity becomes more involved for systems 
involving nonlinear  constraints, as happens, for the motion of a particle on
a $n$-sphere($S^n$). For such systems, Kontsevich
discovered an elaborate graphical rule for the generation of the appropriate
associative star product as a series in ${\hbar}$ \cite{kontsevich}. In this expansion, using 
Cartesian coordinates,  one can see easily that the series deviates from the 
pure exponential form right from $O({\hbar}^2)$ term onwards.
 This indicates that 
for  these systems simple exponentiation of the symplectic bracket 
kernel, having a noncanonical structure, will not yield an associative star product. This is in contrast with 
the above discussed  unconstrained case of a particle moving on a real line
where simple exponentiation of the Poisson bracket kernel yields an associative
star product. 

On the other hand, it has been shown in \cite{batalin} that for any degenerate case, involving only second class
constraints, it is still possible to  express the star product through an exponential factor involving the area of
a geodesic triangle which reduces to a rectilinear triangle, as in the non-degenerate case,
 on the constraint surface. 

At this stage, one can therefore ask whether it is possible to make a suitable 
canonical(point) transformation where the Dirac bracket reduces to the 
usual Poisson bracket in terms of the physical(generalized) coordinates and 
their conjugate momenta with the symplectic matrix $J$ taking the canonical 
form. An answer to this question is provided by a well known theorem of 
Maskawa and Nakajima(MN)\cite{mn} which states that 
for any system described  by a set of canonical variables 
$\Psi_a, \Pi_a ~(a = 1, 2,...N)$ and governed by second class 
constraints, it is possible by a canonical transformation, to construct 
two sets of independent  variables $Q_n, \bar{Q}_r ~(n = 1,2,..M,~ r = M+1,...N)$
 and their respective canonical 
conjugates 
$P_n, \bar{P}_r$, such that the constraints read $\bar{Q}_r \approx 0, 
\bar{P}_r \approx 0$.
Then  the symplectic(Dirac) brackets of the system calculated with respect
to the entire set of variables reduce to Poisson brackets with respect to 
the unconstrained physical variables $Q_n, P_n$ of the system. 
A possible implication of this theorem would be that, when expressed in terms 
of the unconstrained variables, the noncanonical Dirac bracket kernel reduces 
to a canonical Poisson bracket kernel and a simple exponentiation of which  should
then yield an associative star product.  

The objective of this paper is to  study the similarities and dissimilarities of the above mentioned 
non-geometric \cite{groenewold, moyal} and geometric \cite{batalin} approaches through the 
the construction of the MN variables explicitly for a few simple constrained systems.
In this paper we
consider the examples of the Landau problem, particles constrained to move on 
a circle($S^1$)   and a sphere($S^2$). 

The organization of the paper is as follows. Section 2 provides a  review 
of the historical origin, definition and certain properties of star product that
are  relevant to this work. A brief description of MN theorem is also provided. 
Section 3 discusses the Landau problem. Here the constraints are linear and 
it is straightforward to find the canonical transformation that reduces the 
Dirac bracket to the Poisson bracket using the Maskawa-Nakajima theorem. A simple 
exponentiation leads to an associative star product.  Section 4 analyses the 
problem of the motion of a particle on a circle($S^1$) and a sphere($S^2$).
As is well known, here the constraints are nonlinear leading to  variable 
dependent symplectic or Dirac brackets. A naive exponentiation of the 
bracket kernel does not yield an associative star product. A canonical
transformation, which is actually the transition from the Cartesian to the 
spherical polar basis, is employed. Finally, after passing to the 
constraint shell,  we show that the bracket kernel simplifies to a canonical
structure. Now a suitable exponentiation yields an associative star product,
exactly as happens in the unconstrained case. As a bye product of this 
analysis, we get the usual quantization of angular momentum, modulo a half 
factor, in the case of a particle moving on a circle. 
Our conclusions are given in section 5.

\section{Review of Star Product Formalism}
In this section we briefly review the basic ideas of star product formalism
essentially following \cite{zachos, batalin}.
The historical origin of the definition of star product can be traced back to
Groenewold's work \cite{groenewold} wherein it was shown that the space of 
classical phase space functions $f(x, p)$ and the corresponding space of 
Weyl ordered operators $\hat{f}(\hat{x}, \hat{p})$ can have an isomorphism, 
provided the classical 
functions compose through the star product.    
The function 
 $f(x, p)$ is called the classical kernel
of $\hat{f}(\hat{x}, \hat{p})$.
Let us consider the case of a particle moving on a real line ${\cal{R}}^1$ as
an illustrative example.
Clearly the classical phase space($x, p$) is the two dimensional space 
${\cal{R}}^2$.  
An arbitrary phase space function $f(x, p)$ can  be written as
$$
f(x, p) = \int_{-\infty}^{+\infty} dx^{\prime} d p^{\prime} \delta (x - 
x^{\prime}) \delta (p - p^{\prime})  f(x^{\prime} , p^{\prime}) $$
\begin{equation}
 =  \frac{1}{(2\pi)^2}\int_{-\infty}^{+\infty} dx^{\prime} d p^{\prime} 
d\tau d\sigma e^{i[\tau (
x -x^{\prime}) + \sigma (p - p^{\prime})]} f(x^{\prime} , p^{\prime})
\label{1} 
\end{equation}
where the integral representation 
\begin{equation}
\delta (x - x^{\prime}) =  \frac{1}{2\pi} \int_{-\infty}^{+\infty} d\tau
e^{i\tau (x -x^{\prime})}
\label{2}
\end{equation} 
of the Dirac delta function $\delta (x -x^{\prime})$ and a similar 
representation for $ \delta (p - p^{\prime})$ are used. Here $\delta (x - x^{\prime}) $ (\ref{2})
represents the completeness property of the exponential functions $e^{i\tau x}$
(which are eigenfunctions of the Laplacian operator in ${\cal{R}}^1$)  characterized by
 $\tau$ while $\delta (p - p^{\prime})$ represents
the orthonormality.  
In this case, the phase space variables
and hence all the integration variables have noncompact ranges.
At the quantum level, the operator analogues $\hat{x}, \hat{p}$ of $x, p$ 
obey the Heisenberg-Weyl Lie algebra 
\begin{equation}
[\hat{x}, \hat{p}] = i\hbar , \hskip 1.0cm [\hat{x}, \hat{x}] = 0,
\hskip 1.0cm [\hat{p}, \hat{p}] = 0 
\label{3}
\end{equation}
and exp$[i(\tau \hat{x} + \sigma \hat{p} )]$ is a particular element of the
corresponding Lie group.

Weyl's  prescription \cite{weyl}  for arriving at the operator $\hat{f}(\hat{x}, \hat{p})$ 
corresponding to  the kernel $f(x, p)$ (taken to have a polynomial form)  
consists of  rewriting
(\ref{1}) with the replacements $x \rightarrow \hat{x}, p \rightarrow \hat{p}$
to get
\begin{equation} 
\hat{f}(\hat{x},\hat{p}) = \frac{1}{(2\pi)^2}\int_{-\infty}^{+\infty} dx^{\prime} d p^{\prime}
d\tau d\sigma e^{i[\tau (
\hat{x} -x^{\prime}) + \sigma (\hat{p} - p^{\prime})]} 
f(x^{\prime} , p^{\prime}).
\label{4}
\end{equation} 
An equivalent prescription due to Batalin and Tyutin \cite{batalin} is to define\footnote{The choice of the origin $x=p=0$ for evaluating 
$\hat{f}(\hat{x},\hat{p})$ is not mandatory. The operator $\hat{f}(\hat{x},\hat{p})$, evaluated at different points, are in fact related by canonical transformations \cite{batalin}.}
\begin{equation}
\hat{f}(\hat{x},\hat{p}) = e^{[\hat{x} \partial_x + \hat{p}\partial_p ]} f(x, p)|_{x=p=0}
\label{4+1}
\end{equation}
We are however continuing with the prescription (\ref{4}) for the time being.
The quantum operator $\hat{f}$, regarded as a power series in
 $\hat{x}$ and $\hat{p}$,
is first ordered in a completely symmetrized manner using (\ref{3}) and 
a term with $m$ powers of $\hat{x}$ and $n$ powers of  $\hat{p}$ is then
given by the coefficient of $(\tau^m  \sigma^n )$ in the 
expansions of $(\tau \hat{x} + \sigma \hat{p} )^{m+n}$.    

Using the mapping (\ref{4}), Groenewold later  obtained the
classical kernel of the  operator product $\hat{f}\hat{g}$ of two
phase space operators $\hat{f}$ and $\hat{g}$  from the corresponding kernels
 $f$ and $g$ respectively. For that one has to express $\hat{g}(\hat{x}, 
\hat{p})$ just in the manner of $\hat{f}$ in (\ref{4}). One can then write
\begin{eqnarray}
\hat{f}\hat{g}&=&
 \frac{1}{(2\pi)^4} \int {d\xi d\eta d\xi' d\eta'
dx' dx'' dp' dp''}   f(x',p') g(x'',p'')\nonumber \\
&\times  &\exp i(\xi ( {\hat{ p}}-p')+\eta ( {\hat{ x}}-x'))
\exp i(\xi' ( {\hat{ p}}-p'')+\eta' ( {\hat{ x}}-x''))  \qquad
\label{5}
\end{eqnarray}
$$
 =\frac{1}{(2\pi)^4}\!\!\int\! d\xi d\eta d\xi' d\eta'
dx' dx'' dp' dp'' f(x',p') g(x'',p'')
\exp i\left( (\xi +\xi')  {\hat{ p}}+(\eta+\eta')  {\hat {x}}\right)
$$
$$
\times ~\exp i\!\left(-\xi p'-\eta x'-\xi'
p''-\eta'x'' +{\hbar\over 2} (\xi\eta'-\eta\xi') \right).
$$
Changing integration variables to
\begin{equation}
\xi'\equiv {2\over \hbar} (x-x'), \quad
\xi\equiv \tau-  {2\over \hbar} (x-x'), \quad
\eta'\equiv {2\over \hbar} (p'-p), \quad
\eta\equiv \sigma- {2\over \hbar} (p'-p),
\label{6}
\end{equation} 
reduces the above integral to
$$ 
{\hat{ f}}{\hat{g}}= \frac{1}{(2\pi)^2}\int d\tau d\sigma
dx dp \exp i \left(\tau ( {\hat{ p}}-p)+\sigma ( {\hat{x}}-x)\right)$$
\begin{equation} 
\times
\left\{\int dp'
dp''  dx' dx''  ~f(x',p')~g(x'',p'')
 \left[{1\over  (\pi \hbar)^2 }\exp \left(\frac{-2i}{\hbar}
\left( p(x'-x'') + p'(x''-x)+p''(x-x') \right )\right)\right]\right\}.
\label{7}
\end{equation}
In the above equation, consider the exponential inside the square bracket,  
$$
 {1\over  (\pi \hbar)^2 }\exp \left(\frac{-2i}{\hbar}
\left( p(x'-x'') + p'(x''-x)+p''(x-x') \right )\right) $$
$$ ={1\over  (\pi \hbar)^2 } \exp \left(\frac{i}{\hbar}
\left(- 2(p' - p)(x''- x) + 2(x' - x)(p''- p) \right) \right) $$
$$ = \frac{1}{(2\pi)^2}\int d\lambda d\mu \delta (x' - x - {\mu \hbar \over 2}) \delta (p' - p + 
{\lambda \hbar \over 2}) \exp \left({i} \left(\lambda (x''-x) + \mu (p''- p)
\right )\right) $$ 
$$ =\frac{1}{(2\pi)^4} \int d\lambda d\mu d\alpha d\beta \exp i[\alpha(x' - x) + \beta (p' - p)]
\exp {i \hbar \over 2}{(\stackrel{\leftarrow }{\partial}_x \stackrel{\rightarrow }{\partial }_{p}-\stackrel{\leftarrow }{\partial }_{p}
\stackrel{\rightarrow }{\partial }_{x})} \exp \left({i}
 \left(\lambda (x''-x) + \mu (p''- p)
\right )\right) $$
where the representation(\ref{2}) is used. With the aid of the above relation
 one can write the integral in the curly bracket  in (\ref{7}) as,
$$
{1 \over ({2\pi})^4
} \int d\lambda d\mu d\alpha d\beta d x' d p' dx'' dp''  
 \exp i[\alpha(x' - x) + \beta (p' - p)] 
\exp i\hbar{(\stackrel{\leftarrow }{\partial}_x \stackrel{\rightarrow }{\partial }_{p}-\stackrel{\leftarrow }{\partial }_{p}
\stackrel{\rightarrow }{\partial }_{x})/2} $$
$$\times \exp \left({i}
 \left(\lambda (x''-x) + \mu (p''- p)
\right )\right)
~f(x',p')~g(x'',p'')$$
\begin{equation}
=  f(x,p) ~e^{{i \hbar \over 2}(\stackrel{\leftarrow }{\partial }_{x}
\stackrel{\rightarrow }{\partial }_{p}-\stackrel{\leftarrow }{\partial }_{p}
\stackrel{\rightarrow }{\partial }_{x})/2}~ g(x,p).
\label{8}
\end{equation} 
Hence the composition rule is given by, 
\begin{equation} 
{\hat{ f}}{\hat{g}}= \frac{1}{(2\pi)^2}\int d\tau d\sigma
dx dp \exp [i \left(\tau ( {\hat{ p}}-p)+\sigma ( {\hat{x}}-x)
\right)]
(f\star g)(x, p)
\label{9}
\end{equation}
where the  $\star$ product  is defined as,   
\begin{equation}
f(x,p) \star g(x,p) \equiv f(x,p) ~e^{{i\hbar \over 2}(\stackrel{\leftarrow }{\partial }_{x}
\stackrel{\rightarrow }{\partial }_{p}-\stackrel{\leftarrow }{\partial }_{p}
\stackrel{\rightarrow }{\partial }_{x})}~ g(x,p).
\label{10}
\end{equation}
Thus Groenewold \cite{groenewold} showed that $f(x,p) \star g(x,p)$ is the  the classical kernel
 of $\hat{f}\hat{g}$. But this demonstration holds only for systems where particles are moving in 
${\cal{R}}^n$ and corresponds to a non-degenerate case. The same results have been obtained in
\cite{batalin} using a geometrical approach which unlike this derivation does not make use of 
any of the correspondences given in (\ref{4}) or even (\ref{4+1}). i.e., they do not invoke the operators $\hat{x}$ and $\hat{ p}$.
They determine the $\star$ product  just by demanding associativity, i.e., 
\begin{equation}
(f\star g)\star h = f\star(g\star h),
\label{10+1}
\end{equation} 
 neutrality of unity i.e.,
\begin{equation}  
1 \star f = f\star 1 = f
\label{10+2}
\end{equation}
and the classical correspondence
\begin{equation}
\begin{array}{c}
Lim \\
\hbar \rightarrow 0
\end{array} (f\star g) = f.g 
\label{10+3}
\end{equation}
\begin{equation}
\begin{array}{c}  
Lim \\
\hbar \rightarrow 0
\end{array}
 \frac{1}{i\hbar} [f, ~g]_{\star} = \{f,~g\} 
\label{10+4}
\end{equation} 
where $[f, ~g]_{\star} = f\star g -g\star f$ is the Moyal bracket. Eq (\ref{10+3}) clearly demonstrates that $\star$ multiplication is  
really a deformation of the ordinary multiplication. The latter is restored when the deformation parameter $\hbar \rightarrow 0$. 
Since the star product involves exponentials of derivative operators, it is 
possible in certain cases to write it as a translation in function arguments;
\begin{equation}
f(x,p) \star g(x,p) = f\left( x+{i\hbar\over 2}\rp_p ,~ p-{i\hbar\over 2}\rp_x
\right )~ g(x,p)
\label{11}
\end{equation}
so that, as a corollary, we get the Moyal brackets 
\begin{equation}  
[x, p]_{\star}  = i\hbar, \hskip 1.0cm
 [x, x]_{\star} = [p, p]_{\star} = 0 
\label{12}
\end{equation}
which is the counterpart of the Heisenberg-Weyl Lie algebra (\ref{3}). Clearly these Moyal brackets trivially satisfy the conditions 
(\ref{10+3}) and (\ref{10+4}).
Conversely, using the defining relation (\ref{10}) 
  it is easy to cross-check algebraically that the star product is 
indeed associative owing to the form of the exponential involved in the relation. 
Although the ranges of the phase space variables $(x, p)$ is $(-\infty , +\infty )$, this algebraic demonstration {\it does not} depend
on this. 
The associativity of the star product in this case 
(1-dimensional free particle) also follows from a geometrical  approach\cite{zachos}. The expression multiplying ($-2i$/$\hbar$) in the second exponential 
of (\ref{7}) can be written 
as 
\begin{equation}
2A({\bf r''}, {\bf r'}, {\bf r} ) = ({\bf r}' -{\bf r})\times
 ({\bf r}-{\bf r}'')=
{\bf r}''\times {\bf r}'   +
{\bf r}'\times {\bf r}+
{\bf r}\times {\bf r}'' 
\label{13}
\end{equation}
where $A({\bf r''}, {\bf r'}, {\bf r}) $ is the area of the phase space 
triangle $({\bf r}'',{\bf r}',{\bf r})$, with  ${\bf r}\equiv (x,p)$.
Thus $f\star g$ can be expressed in a more compact form as, 
\begin{equation} 
f\star g({\bf r}) = {1\over  (\pi \hbar)^2 } \int d{\bf r'}d{\bf r''} 
  ~f({\bf r}')~g({\bf r}'')\exp \left(\frac{4}{i\hbar}A({\bf r''}, {\bf r'}, {\bf r} )\right).
\label{1-11}
\end{equation}

This same form for the exponent involving the area of a triangle in phase space was obtained
in \cite{batalin} in a coordinate independent manner. There it involved the integral of a 
closed symplectic 2-form 
\begin{equation}
J = {1 \over 2} J_{ij} dz^i \wedge dz^j; ~~~~~dJ = 0 
\label{1-111}
\end{equation} 
along a surface spanned by the contour of a geodesic triangle in a flat (i.e., vanishing Riemann tensor\footnote{Note that the 
phase space is not endowed with any metric as such.}) even-dimensional phase space 
having $z^i$ as coordinates. The sides of the geodesic triangle are determined 
relative to the symmetric connection which provides a covariant constancy of the symplectic 2-form. As a matter of notation
the matrix $\{J^{ij}\}$ - the inverse of the matrix $\{J_{ij}\}$ occurring in (\ref{1-111}):
\begin{equation}
J^{ij} J_{jk} = \delta^i_k
\label{1-1111}
\end{equation}
is used to define Poisson bracket as 
\begin{equation}
\{f, ~g\} = f \Lambda g \equiv f \left(\sum_{i, j} \lp_i J^{ij} \rp_j \right) g
\label{1-11111}
\end{equation}
and $\Lambda$ is referred to as the Poisson bracket kernel. Since the phase space is taken to be flat,
Darboux theorem   ensures the existence of a suitable coordinate system, where $\{J_{ij}\}$
becomes a constant matrix and the connection vanishes so that the above mentioned geodesic 
triangle reduces to a rectilinear triangle. By suitable transformation involving scaling etc, this matrix can be 
further reduced to the following canonical form
\begin{equation} 
\{J_{ij}\} = 
 \left(
\begin{array}{cc}
        0 & I \\
        -I & 0
        \end{array} \right).
\label{1-111111}
\end{equation}
Note that the star product defined in (\ref{10}) is nothing but the exponential  of 
the Poisson bracket kernel $\Lambda$ sandwiched between the functions $f$ and $g$ 
with the matrix $\{J_{ij}\}$ given in (\ref{1-111111}).

Now the associativity property of the star product can  be easily seen from this
representation because of the appearance of the area of a phase space 
triangle in the exponent. 
For triple star product involving functions $f, g$ and $h$ we have
\begin{equation}
(f\star g) \star h={1\over \hbar ^4 \pi^4}\!\int\! d\overline{\bf r} d{\bf r}'
 d{\bf r}'' d{\bf r}'''f({\bf r}') g({\bf r}'') h({\bf r}''') 
\exp \left[{4\over i\hbar} \left(
A({\bf r}'', {\bf r}',\overline{\bf r})+A({\bf r}''', \overline{\bf r}, {\bf r})
 \right)\right].
\label{1-12}
\end{equation}
The $d\overline{\bf r}$ integrations can be exploited to obtain
corresponding $\delta$-functions and the product can be simplified to
$$
((f\star   g)  \star h)  ({\bf r}) =
{1\over \hbar ^2 \pi^2}\! \int \! d{\bf r}' d{\bf r}'' d{\bf r}''' 
f({\bf r}') g({\bf r}'') h({\bf r}''')
$$
\begin{equation}
\times  \delta ({\bf r}-{\bf r}'+{\bf r}''-{\bf r}''') 
~\exp \left({4\over i\hbar} A({\bf r}''',{\bf r}'',{\bf r}', {\bf r}) \right)    .
\label{1-13}
\end{equation}
The argument of the delta functions in the above expression ensures that
$ A({\bf r}''', {\bf r}'',{\bf r}', {\bf r})$ is the area of the parallelogram
 with vertices 
$({\bf r}''',  {\bf r}'',{\bf r}',{\bf r})$. 
Therefore, the order
in which the functions $f, g$ and $h$ are associated becomes immaterial thus
providing pictorial proof of the associativity. 

The Groenewold derivation and the area interpretation following from it,  as discussed above
relied crucially on the Fourier expansion (\ref{1}) 
and is related to the fact that the simple exponential functions are 
eigenfunctions of the Laplacian operator $(\frac{\partial^2}{\partial x^2})$
 in ${\cal R}^1$. These exponential functions provide a complete orthonormal 
basis for the space of functions defined on ${\cal R}^1$.
Therefore,
this  may not necessarily be implementable in the case of physical
systems where the eigenfunctions of the Laplacian operator, although form a
 complete set of basis, are not given by simple exponential functions.  
For example, in the case of a particle constrained to $S^2$, to be discussed subsequently, 
the expansion corresponding to (\ref{1}) should be obtained from an 
appropriate harmonic analysis where the spherical harmonics 
$Y_{lm}(\theta, \phi)$ are the  
eigenfunctions of the Laplacian operator and
 provide the complete set of basis. 
Here the $\theta$ dependence is not given by the exponential functions anymore.
Nevertheless, the geometrical approach of Batalin and Tyutin \cite{batalin} shows that here too 
one gets a geodesic triangle in the exponent of the star product, thus demonstrating the superiority of 
the geometrical approach.

It is also very clear from the above analysis that the star product between classical phase space functions incorporates some
of the quantum features right at the classical level. This gives rise to the possibility that quantum mechanics can be formulated
in terms of the classical phase space variables \cite{zachos}. This is because the classical phase functions must compose, as we
have seen, through the star product in order to have an isomorphism between the set of classical phase space functions and
Weyl ordered operators.
We shall see another beautiful demonstration of this
in section 4, where we will be dealing with a particle moving on a circle.

Coming to the degenerate case, i.e., for systems with nonlinear second class 
constraints the star product
is to be obtained by the exponentiation of the Dirac bracket kernel as has been shown in \cite{batalin}.
But unlike the Poisson bracket kernel, the Dirac bracket kernel may not in general lead to
a manifestly associative star product if Cartesian coordinates are used. 
This is because the $J_{ij}$ are now dependent on the phase space variables.  
To circumvent this difficulty, Kontsevich\cite{kontsevich} devised detailed graphical rules
wherein  the star product can be expressed as a series expansion in $\hbar$.
$$ f\star g = f [
1+{i\hbar \lp_i J_{ij} \rp_j}
-\frac{\hbar^2}{2}\left (  \lp_i \lp_k J_{ij} J_{kl} \rp_j \rp_l \right)
\qquad \qquad \qquad \qquad
$$
\begin{equation}
-\frac{\hbar^2}{3}\left (  \lp_i \lp_k J_{ij} (\partial_j J_{kl}) \rp_l
- \lp_k J_{ij} (\partial_j J_{kl}) \rp_i \rp_l  \right)
+O(\hbar^3)] g.
\label{16}
\end{equation}
One can see that the series deviates from the exponential  right from
the $O(\hbar^2)$ term onwards. However associativity is restored.
One may also notice that if the $J_{ij}$'s are constants(as in the case of
the Poisson bracket kernel in Darboux coordinates), the extra terms vanish thereby restoring 
the exponential form of the series. Associativity of the star product is then
automatically guaranteed.

It is here that the  Maskawa-Nakajima theorem \cite{mn}
provides a clue as to how one can obtain a manifestly associative star product 
for  second class constraint systems without using the Kontsevich series
because of the possibility of reducing the Dirac bracket kernel to Poisson 
bracket kernel by MN's appropriate canonical transformations.
 We now give a brief outline of the method \cite{weinberg}. 
According to Dirac's prescription\cite{dirac}, for a system with second class constraints
the Poisson bracket of two phase space functions $A$ and $B$
 should be generalized to the Dirac bracket(DB):
\begin{equation}
\{A, B\}_{DB} = \{A, B\} - \{A, \Omega_i\} (C^{-1})_{ij} \{\Omega_j, B\} 
\label{17}
\end{equation}
where $\Omega_i \approx 0$ are the constraints of the system and
\begin{equation}
C_{ij} \equiv \{\Omega_i, \Omega_j \}
\label{18}
\end{equation}
is the matrix of the Poisson brackets of the constraints.
The Maskawa-Nakajima theorem\cite{mn} states that for a second class constraint system
there exists a set of independent canonical variables 
$(Q_n, \bar{Q}_r, P_n, \bar{P}_r)$
where $P_n, \bar{P}_r$ are canonical conjugates of $Q_n, \bar{Q}_r$ 
respectively,
such that the constraints of the system are 
\begin{equation}
\Omega_{1r} =  \bar{Q}_r \approx 0, 
\hskip 1.0cm \Omega_{2r} =  \bar{P}_r
\approx 0
\label{19}
\end{equation}
After the rearrangement, the elements of the matrix $C$ read
\begin{equation} 
C_{1r, 2s} = \{\bar{Q}_r, \bar{P}_s \} = \delta_{rs} \hskip 1.0cm C_{1r, 1s} = 
\{\bar{Q}_r, \bar{Q}_s \} = 0 \hskip 1.0cm  C_{2r,  2s} = \{\bar{P}_r, 
\bar{P}_s \} = 0
\label{20} 
\end{equation} 
and Poisson brackets of any function $A$ with the constraints now become
\begin{equation} 
\{A, \Omega_{1r} \} = -\frac{\partial A}{\partial \bar{P}_r}, \hskip 1.0cm 
\{A, \Omega_{2r} \} = \frac{\partial A}{\partial \bar{Q}_r}
\label{21}
\end{equation} 
Since the $C$ matrix has the property $C^{-1} = -C$, the Dirac bracket can be 
written as 
$$
\{A, B\}_{DB} = \{A, B\} + \{A, \Omega_{1r} \} \{\Omega_{2r}, B \} - \{A, \Omega_{2r}
 \} \{\Omega_{1r} , B \} $$
\begin{eqnarray} 
& = & \{A,  B\} - \frac{\partial A}{\partial \bar{Q}_r}  \frac{\partial B}
{\partial \bar{P}_r}  + \frac{\partial B}{\partial \bar{Q}_r} \frac{\partial A}
{\partial \bar{P}_r} \\
& = & \frac{\partial A}{\partial {Q}_n}  \frac{\partial B}
{\partial {P}_n}  - \frac{\partial B}{\partial {Q}_n} \frac{\partial A}
{\partial {P}_n}
\label{22}
\end{eqnarray} 
That is, the Dirac bracket is equal to a Poisson bracket computed 
with respect to the reduced set of unconstrained physical variables 
$(Q_n, P_n)$. 
Since the Poisson bracket kernel  can be exponentiated to yield an associative 
star product, the problem of constructing a manifestly associative star product for a 
second class system  reduces to that of finding out the correct set of 
canonical variables in terms of which the Dirac brackets become equal to 
Poisson brackets. The existence of such canonical 
variables is guaranteed by the Maskawa-Nakajima theorem though the actual task of obtaining
those variables for a given system may be highly nontrivial. Once the
proper variable have been identified the relevant bracket kernel, after
passing to the constraint shell, assumes the standard form with the 
$J$  becoming a constant symplectic matrix. Associativity 
of the star product then follows naturally.

Here we would like to mention that the Maskawa-Nakajima theorem, strictly speaking,
ensures the existence of only a local coordinate system that separates constrained and unconstrained
variables. But the Fourier transform reflects the global topology, as we have discussed earlier with the example of 
$S^2$.  As we shall see later (in section 4.3), the polar coordinates $(\theta , \phi)$ can be identified with the
physical variables for $S^2$. Clearly these variables are not defined globally, as the north and south poles have to 
be excluded to define these coordinates unambiguously. In fact, a single coordinate chart which covers the entire $S^2$
does not exist at all. Also the choice of spherical harmonics $Y_{lm}(\theta, \phi )$ as the basis of expansion, satisfying 
both orthonormality and closure relations, although reflects the global topology of the base manifold $S^2$, can only be defined in a coordinate neighborhood. 

However the absence of a single coordinate chart covering the entire $S^2$ is not a serious problem, as one can construct an 
atlas comprising of several compatible charts covering the entire $S^2$. Since the transition from one chart to another
amounts to a point canonical transformation, it is clear that the final Poisson bracket kernel will remain invariant under 
such a transformation.  We show this through a simple example in section 4.1,
 where the canonical form of the bracket kernel (in ${\cal S}^1$) is preserved as one makes a transition
 from polar to stereographically 
projected coordinates. Besides, the algebraic demonstration of associativity 
of the star product for a degenerate (second class constrained) system through order by order (in $\hbar$) expansion 
becomes exactly similar to the non-degenerate case once the Dirac bracket kernel is reduced to the Poisson bracket kernel
through the identification of the physical variables, whose existence is ensured by MN theorem. The curvilinear nature of these 
coordinates then become completely inconsequential. Also one does not need to invoke the ``triangle argument". 

For simple  second class systems like the Landau problem, a particle 
constrained to move along a circle or on the surface of a sphere etc.,
it is possible to obtain the exact form of these unconstrained physical 
variables from symmetry considerations. Usually, star products even for 
these systems are defined in a complicated manner using the Kontsevich series
expansion method.   In the subsequent sections we  demonstrate
with the help of these systems, how one can arrive at a manifestly  associative star
product, although their Dirac brackets in the Cartesian system do not reveal this.

\section{Landau Problem}

The Lagrangian for the classical Landau problem of a spinless 
charged  particle moving on a two dimensional  plane, in a constant 
background magnetic field  $B$, can be written as,
\begin{equation}
L=\frac{m}{2}\dot{\bf{x}}^2 +  \frac{B}{2} {\bf x} \times  \dot{\bf x} - \frac{k}{2} {\bf x}^2,
\label{23}
\end{equation}
where a harmonic oscillator potential is chosen. 
When the magnetic field is very
strong we can suppress the kinetic term(which is equivalent to the limit $m 
\rightarrow 0$) and the Lagrangian then describes a first order constrained 
system. 
Writing the kinetic term in the component form, we get, 
\begin{equation}
L= \frac{B}{2} x_i \epsilon_{ij} \frac{dx_j}{dt} -
\frac{k}{2} {\bf x}^2
\label{24}
\end{equation} 
The canonically conjugate  momenta are constrained to the respective transverse
coordinates,
\begin{equation}
\Omega_i = p_i + \frac{B}{2} \epsilon_{ij} x_j \approx 0
\label{25}
 \end{equation}
giving  the  two second class constraints.
The corresponding Dirac brackets are given by 
\begin{equation}
\{ x_i , x_j\}_{DB} =   -\frac{\epsilon _{ij}}{B}~, \qquad 
\{  x_i, p_j  \}_{DB}   =\frac{\delta_{ij}}{2}, \qquad
\{  p_i, p_j   \}_{DB}  =-\frac{B}{4} \epsilon_{ij} .
\label{26}
\end{equation}
Thus the  two perpendicular directions $ x_1$ and $x_2 $ do not 
commute(in the sense of classical
Dirac brackets),
rather they behave as canonical conjugates to each other.
The Dirac bracket for two phase space functions $f$ and $g$ can be written as 
$$
\{ f,g \}_{DB} = f(\lp_{x_i} \frac{1}{2} \rp_{p_i}
-\lp_{p_i} \frac{1}{2} \rp_{x_i} -\lp_{x_i} \frac{\epsilon_{ij}}{B} \rp_{x_j}
-\lp_{p_i} \frac{\epsilon_{ij} B}{4} \rp_{p_j}) g \equiv f \Lambda g 
$$
with 
\begin{equation} 
f \Lambda g =f \lp_{\mu}  ~J^{\mu \nu}~ \rp_{\nu} g ~.
\label{28}
 \end{equation}
where  the symplectic matrix is now given by
\begin{equation}
\{J^{\mu \nu}\} = \left( \begin{array}{cccc}
0 & -\frac{1}{B} & \frac{1}{2} & 0 \\
\frac{1}{B} & 0 & 0 & \frac{1}{2} \\
-\frac{1}{2} &  0 & 0 & -\frac{B}{4} \\
0 & -\frac{1}{2} & \frac{B}{4} & 0 
\end{array} \right).
\label{29}
\end{equation}
In the example considered, the space where the antisymmetric matrix $\{J^{\mu \nu}\}, ~ (\mu, \nu = 1,..,4)$
acts is 4-dimensional  (in the order $x_1, x_2, p_1, p_2$). Although $\{J^{\mu \nu}\}$ does not have the canonical form 
(\ref{1-111111}) the exponentiation
 of this Dirac bracket kernel gives an associative star product  as one can see 
immediately. This is because the $J$ matrix is a constant matrix arising
from the linear form of the constraints (\ref{25}). The corresponding phase space
coordinates $(x_i, p_j)$ can therefore already be identified with Darboux coordinates. This, however,
 will not be true for systems involving nonlinear constraints. There one has 
to isolate the physical degrees of freedom, following a systematic technique. 

To see the power of this technique of working with
the physical variables, we apply it to the present case and show that the
associativity becomes  even more transparent, in the sense that 
$\{J^{\mu \nu}\}$ matrix (\ref{29}) takes the canonical form (\ref{1-111111}) and
it becomes 
similar to an unconstrained system.   
To that end,  make a canonical transformation from
$(x_1, x_2, p_1, p_2)$ variables to a new set of variables given by 
$(\bar{X}, X, \bar{P}, P)$,  where $(X, P)$ and $(\bar{X},\bar{P})$ are
the new independent canonical pairs,
\begin{eqnarray}    
\bar{X} & = & \frac{p_1}{2} +  \frac{B}{4}x_2 \\
\bar{P} & = & \frac{2}{B}p_2 - x_1 \\
X  & = & x_1 + \frac{2}{B}p_2 \\
P  & = &\frac{p_1}{2} - \frac{B}{4}x_2.
\label{30}
\end{eqnarray}
It may be noted that the new variables $\bar{X}, \bar{P}$ are 
proportional to the constraints (\ref{25}) of the system 
$(\bar{X} = \frac{1}{2}\Omega_1,~~
\bar{P} = \frac{1}{2}\Omega_2)$ while the physical variables are 
$X$ and $P$. Its implication for the bracket kernel is now discussed.
In terms of the new  basis $(\frac{\partial}{\partial \bar{ X}}, \frac{\partial}{\partial  X},
\frac{\partial}{\partial \bar{ P}}, \frac{ \partial}{\partial  P})$ for the tangent space of the phase 
space, 
the old ones are given  by
\begin{eqnarray*}
\frac{\partial}{\partial x_1} & = &   \frac{\partial}{\partial X}  - \frac{\partial}{\partial \bar{P}}   \\
\frac{ \partial}{\partial  x_2} & = & \frac{B}{4}\left(\frac{\partial}{\partial 
\bar{X}} - \frac{\partial}{\partial P} \right) \\  
\frac{ \partial}{\partial  p_1 } & = & \frac{1}{2}\left(\frac{\partial}{\partial
\bar{X}} + \frac{\partial}{\partial P} \right) \\ 
\frac{ \partial}{\partial  p_2} & = & \frac{2}{B}\left( \frac{\partial}{\partial X} + \frac{\partial}{\partial
\bar{P}} \right). \\  
\end{eqnarray*} 
The bracket kernel, in the new variables, is given by
\begin{equation}
\Lambda =~~ \lp_{\mu} J^{\mu \nu} \rp_{\nu}~~ =~~ \lp_X \rp_P - \lp_P \rp_X
\label{31}
\end{equation}
which is the standard canonical form. Hence the associativity of the star 
product follows trivially, being exactly identical to the unconstrained case.
Note that the 4-dimensional phase space where the original kernel was defined
reduces to the physical two dimensional space, when using the new variables.
The two dimensional constraint sector just cancels out. 
\section{Particle on $n$-Sphere $(S^n)$}
The Lagrangian of a particle restricted to move on the surface $S^n$
of unit radius is 
given by 
\begin{equation}
L = \frac{1}{2}(\dot{\bf x}^2) -\lambda ({\bf x}^2 - 1) \hskip 1.0cm \bf x 
\in {\cal R}^{n+1}
\label{32}
\end{equation} 
where $\lambda$ is a Lagrange multiplier
 enforcing the primary constraint
\begin{equation}
\Omega_1 = {\bf x}^2 - 1 \approx 0.
\label{33}
\end{equation}
Time conservation of the primary constraint,
\begin{equation} 
\{ \Omega_1, H_T \} = 0 
\label{34}
\end{equation}
leads to the secondary constraint
\begin{equation}
\Omega_2 = {\bf x}\cdot {\bf p} \approx 0
\label{35}
\end{equation}
with the total Hamiltonian $H_T$ being given by
\begin{equation}
 H_T =  H_c + \lambda \Omega_1
\label{36}
\end{equation}
where $H_c$ is the canonical Hamiltonian obtained from (\ref{32}).
The pair $\Omega_1, \Omega_2$ form a second class system of constraints.
The corresponding 
Dirac brackets in component form are
\begin{equation}
\{x_i, x_j\}_{DB} = 0, \hskip 1.0cm \{x_i, p_j\}_{DB} = \delta_{ij} - x_ix_j,
\hskip 1.0cm \{p_i, p_j\}_{DB} = x_jp_i - x_ip_j,
\label{37}
\end{equation}
and the Dirac bracket kernel is, 
\begin{equation}
\Lambda = \lp_x\!\cdot\!\rp_p - \lp_x\!\cdot x  x\cdot\!\rp_p -\lp_p\!\cdot\rp_x
+\lp_p\!\cdot x x\cdot \rp_x + \lp_p\!\cdot p x\cdot\!\rp_p
-\lp_p\!\cdot x p\cdot\!\rp_p.
\label{38}
\end{equation}
The exponentiation of this kernel does not yield a manifestly associative star product. 
However, the Maskawa-Nakajima theorem helps us to write it in a manifestly associative form. We illustrate this
first with the example of a particle constrained to move along a circle 
$S^1$ and then
in the case of a particle on a 2-sphere $S^2$.
\subsection{Reduction of the Bracket Kernel for $S^1$ and Associativity}
The simplest case of a phase space with the Dirac brackets (\ref{37}) 
is that of particle moving along a unit circle described by 
(\ref{32}) with $ i = 1, 2$. The symplectic matrix associated with the 
bracket kernel of this system is given by
\begin{equation}
\{J^{\mu \nu}\} = \left( \begin{array}{cccc}
0 & 0 & 1-x_1^2 & -x_1x_2 \\
0 & 0 & -x_1x_2 & 1- x_2^2 \\
-(1-x_1^2) & x_1x_2 & 0 & x_2p_1 -x_1p_2 \\
x_1x_2 & -(1-x_2^2) & -(x_2p_1 -x_1p_2) & 0 
\end{array} \right)
\label{39}
\end{equation}
which acts on the phase space spanned by the coordinates $(x_1, x_2, p_1, p_2)$.
    
We now consider 
the following new set of phase space coordinates which can be obtained from
the set $(x_1, x_2, p_1, p_2)$ by a canonical transformation;
\begin{eqnarray}
\bar{x} & = & \frac{1}{\sqrt{2} } (x_i^2 - 1) \\
\theta & = & \tan^{-1} \frac{x_2}{x_1} \\
\bar{p} & = & \frac{1}{\sqrt{2} } x_i p_i  \\
p_{\theta} & = & x_1p_2 -x_2p_1 
\label{40}
\end{eqnarray}
where $(\bar{x}, \bar{p})$ and $(\theta , p_{\theta})$ are the new independent canonical pairs. Observe that the canonical pair $(\bar{x}$ and $\bar{p})$
corresponds respectively to the constraints (\ref{33}) and (\ref{35}) 
of the system and normalized suitably to yield $\{\bar{x}, \bar{p}\} = 1$. 
Also note that $\theta$ is the angle coordinate and $p_\theta$ is the 
angular momentum. Indeed, the above canonical transformation is the usual 
transformation that maps from a Cartesian to polar basis. Correspondingly, the tangent space basis of the phase
space $(\frac{\partial }{\partial x_i}, \frac{\partial }{\partial p_i})$ undergo
a suitable transformation. The old tangent space basis in terms of the new one 
is given by  
\begin{eqnarray*} 
\frac{\partial}{\partial x_1} & = & \sqrt{2}  \cos \theta ~\frac{\partial}{\partial\bar{x}}
~ +~ \frac{1}{\sqrt{2}} \left( \sqrt{2} \cos \theta ~\bar{p} - \sin \theta~  
p_{\theta} \right) \frac{\partial}{\partial\bar{p}} \\
& = & ~ -~ \sin\theta ~\frac{\partial}{\partial \theta}~ +~
\left(\cos \theta ~p_{\theta} +  \sqrt{2} \sin\theta ~ \bar{p}\right)  
\frac{\partial}{\partial p_{\theta}} \\
\frac{\partial}{\partial x_2}  & = & \sqrt{2}  \sin\theta ~\frac{\partial}{\partial\bar{x}}
~ +~ \frac{1}{\sqrt{2}} \left(\cos \theta ~p_{\theta} +  \sqrt{2} \sin\theta~ \bar{p}\right)~\frac{\partial}{\partial \bar{p}} \\
& = & ~ + ~ \cos \theta ~\frac{\partial}{\partial \theta}~ - ~
\left( \sqrt{2} \cos \theta ~\bar{p} - \sin \theta~
p_{\theta} \right)~\frac{\partial}{\partial p_{\theta}} \\ 
\frac{\partial}{\partial p_1} & = & \frac{\cos\theta}{\sqrt{2}} 
\frac{\partial}{\partial\bar{p}} ~-~ \sin \theta 
 \frac{\partial}{\partial p_{\theta}} \\
\frac{\partial}{\partial p_2} & = & \frac{\sin\theta}{\sqrt{2}}
\frac{\partial}{\partial\bar{p}} ~+~ \cos \theta
 \frac{\partial}{\partial p_{\theta}} 
\end{eqnarray*} 
Making use of the above set of relations one can show that the bracket kernel
(\ref{38}) can be expressed as,
\begin{equation}
\Lambda = \lp_{\theta} \rp_{p_{\theta}} - \lp_{p_{\theta}} \rp_{\theta}
+\frac{\bar{x}}{\sqrt{2}} (\sqrt{2}\bar{x} + 1) \left(\lp_{\bar x} \rp_{\bar p} - \lp_{\bar p}\rp_{\bar x}
\right)
\label{42}  
\end{equation}
If we now pass to the constraint shell $\bar{x}  \approx 0$, the second term in (\ref{42})
 vanishes so that 
\begin{equation}
\Lambda = \lp_{\theta} \rp_{p_{\theta}} - \lp_{p_{\theta}} \rp_{\theta}
\label{43}
\end{equation}
This is the standard canonical form. In contrast to the Landau problem, the 
constraint part does not simply cancel by itself. It is necessary to pass to
 the constraint shell at the end of the computation to obtain the desired 
canonical structure for $\Lambda$. 
Using this bracket kernel one defines \cite{batalin} the star product as an exponential series
in $\hbar$ just as  ({\ref{10}) to yield
\begin{equation} 
 f(\theta, p_{\theta}) \star g(\theta, p_{\theta}) = f(\theta, p_{\theta}) 
\exp\left[ \frac{i\hbar}{2}\left(\lp_{\theta} \rp_{p_{\theta}} - 
\lp_{p_{\theta}} \rp_{\theta}\right) \right] g(\theta, p_{\theta})
\label{44}
\end{equation} 
Clearly this has the same form as (\ref{10}) except that the $\theta$ variable 
occurring here has a compact range $(0\leq \theta < 2\pi)$. As far as 
algebraic demonstration  of the associativity is concerned, this is
inconsequential. Associativity is demonstrated exactly as happens in the 
unconstrained case. 

If one uses a coordinate system different from $\theta$, then also the Dirac bracket 
kernel takes the canonical form of the Poisson bracket kernel.
An illustrative example is the stereographic projection of $S^1$ to $R^1$ given by,
\begin{equation}
{\cal X} = \frac{\cos \theta}{1 - \sin \theta}.
\label{46+1}
\end{equation}
The momentum conjugate to ${\cal X}$ is 
\begin{equation}
{\cal P}_{\cal X} = (1 - \sin \theta) p_\theta.
\label{46+2}
\end{equation}
The transition from $(\theta, p_\theta)$ to $({\cal X}, {\cal P})$ is a 
canonical transformation and  
it is easy to see that, in terms of these new variables ((\ref{46+1}) and (\ref{46+2})),  
 the kernel (\ref{43}) takes the canonical form 
\begin{equation}
\Lambda~~ = ~~\lp_{\cal X} \rp_{\cal P_{\cal X}} - \lp_{\cal P_{\cal X}} \rp_{\cal X}
\label{pcbk}
\end{equation}
which obviously  gives a manifest associativity for the star product.
\subsection{Geometrical Proof of Associativity and
 Angular Momentum Quantization}

The associativity of the star product in the preceding example  can  also 
be seen  from the triangle interpretation.
However, in the present case,  the  coordinate 
variable $\theta$  has the range
 $0 \le \theta < 2\pi$ while its conjugate momentum varies 
continuously(classically) 
over a noncompact
range $(-\infty, +\infty)$. 
Hence the classical phase space of the system has
 the geometry of an infinite cylinder of unit radius.
Therefore, one has to make appropriate modifications in the
Fourier representation of $f(\theta, p_{\theta})$ in order to verify 
the associativity
using the triangle interpretation. 
That is,  one writes the  phase space function $f$ as
$$
f(\theta, p_{\theta}) =  \int_{-\pi}^{+\pi} d\theta' \int_{-\infty}^{+\infty} 
dp' \delta_p(\theta - \theta')\delta (p_{\theta} - p') f(\theta' , p') $$
\begin{eqnarray}
& = & \frac{1}{(2\pi )^2} \sum_{n} \int_{-\pi}^{+\pi} d\theta' 
\int_{-\infty}^{+\infty} dp' d\phi e^{i[n(\theta - \theta ') + \phi (p_{\theta} - p')]}
f(\theta ', p').
\label{47}
\end{eqnarray}
where $n$ takes integer values and 
\begin{equation}
\delta_p(\theta - \theta') ={1 \over 2\pi}
\sum_n e^{in(\theta - \theta')} 
\label{48}
\end{equation}
is the periodic delta function.\footnote{Notice that  (\ref{48})
is the completeness relation for the exponential functions $e^{in\theta}, ~
(n \in {\cal Z})$ which provide a  complete set of basis in the space 
of functions defined on $S^1$. Also, any function 
$f(\theta)$ defined on $S^1$ satisfies the usual property of a delta function, viz., $f(\theta) = \int_{-\pi}^\pi d\theta ' \delta_p(\theta - \theta') f(\theta ')$.} 
Note that the periodic delta function is 
incorporated on account of the periodicity of $\theta$ whereas the usual 
Dirac delta function suffices in the case of $p_\theta$ which has a noncompact
range.
Therefore the star product of two functions $f(\theta, p_{\theta})$ and 
$g(\theta, p_{\theta})$ using this Fourier
representation is to be written as
$$f \star g = 
\frac{1}{(2\pi )^4}\sum_{m, n}\int_{-\pi}^{+\pi} d\theta ' d \theta '' 
\int_{-\infty}^{+\infty} dp' dp'' d\phi d\chi  $$
\begin{equation} \times
e^{i[n(\theta - \theta ') +
 \phi (p_{\theta} -p')]} e^{\frac{i\hbar}{2}\left[\lp_{\theta} 
\rp_{p_{\theta}} - 
 \lp_{p_{\theta}} \rp_{\theta}\right]} e^{i[m(\theta - \theta '') +
 \chi (p_{\theta} - p'')]} f(\theta', p') g (\theta '', p''). 
\label{49}
\end{equation}
Note at this stage that the c-number valued exponential functions appearing on 
either side of the $\star$ operator are eigenfunctions of 
$\partial_{\theta}$ and$\partial_{p_\theta}$. Therefore, one can combine 
all these exponential factors into a single one and a subsequent integration
over $\phi$ and summation over $n$ yields,
$$f \star g = \frac{1}{(2\pi )^4}\sum_{m, n}\int_{-\pi}^{+\pi} d\theta ' d \theta ''
\int_{-\infty}^{+\infty} dp' dp'' d\phi d\chi $$
$$\times  e^{i[n(\theta - \theta ' -
\frac{\hbar \chi}{2} ) + \phi (p_{\theta} -p'+ \frac{\hbar m}{2} ) +
\chi (p_{\theta} - p'') + m(\theta - \theta '')]}  f(\theta', p')
 g (\theta '', p'') 
$$
$$ = \frac{1}{(2\pi )^2}\sum_{m}\int_{-\pi}^{+\pi} d\theta ' d \theta ''
\int_{-\infty}^{+\infty} dp' dp''  d\chi \delta (p_{\theta} - p' + \frac{\hbar m}{2} ) 
 \delta_p(\theta - \theta ' - \frac{\hbar \chi}{2}  ) $$
 $$\times
e^{i[m(\theta - \theta '') + 
 \chi (p_{\theta} -p'')]} f(\theta ', p') g (\theta '', p'').
$$
This involves a pair of delta functions,
one of which is periodic. We next perform the integration over $\chi$
and use the properties of the periodic  delta functions\footnote{For any 
function $f(x)$ defined over the real line $R^1$, the periodic delta function 
satisfies $\int_{-\infty}^{+\infty} dx \delta_p(\theta - x) f(x) = 
\sum_n e^{in\theta} \tilde{f}(n)$ where $\tilde{f}(n) = \tilde{f}(k)|_{k= n}$
 and  $\tilde{f}(k) = \frac{1}{2\pi} \int_{-\infty}^{+\infty} dx e^{-ikx} f(x)$
is the Fourier transform of $f(x)$.} to rewrite
the integral in terms of the ordinary (non periodic) Dirac delta functions
as follows;
$$f\star g = \frac{1}{(2\pi )^2}\sum_{m, n} \int_{-\pi}^{+\pi} d\theta '
 d \theta '' \int_{-\infty}^{+\infty} dp' dp'' $$ 
\begin{equation}
\times \delta (p_{\theta} - p' + 
\frac{\hbar m}{2} ) \delta(p_{\theta} - p'' - \frac{\hbar n}{2}  ) 
e^{\frac{2i}{\hbar}[-(p_{\theta} - p') (\theta - \theta '') + (p_{\theta} -p'')
(\theta - \theta ')]}f(\theta ', p') g (\theta '', p'').  
\label{50}
\end{equation} 
Here we notice the presence of a set of delta functions in the integral which
restrict the difference between any pair among the variables
 $p_{\theta}, p', p''$ to integral multiples of $\frac{\hbar}{2}$. 
\begin{equation}
p ' -  p_{\theta}= \frac{\hbar n_1}{2}, \hskip 1.0cm 
p'' -  p_{\theta} =  \frac{\hbar n_2}{2} \hskip 1.0cm
p ' - p'' = \frac{\hbar n_3}{2} 
\label{51}
\end{equation} 
where $n_1,n_2,n_3 $ are some integers. General solutions to the above set 
of three equations,
consistent with the  chiral symmetry of the system(clockwise 
 $\leftrightarrow$  anticlockwise  interchange in this case), are given by, 
\begin{equation}
p_{\theta} = \frac{\hbar m_1}{2}, \hskip 1.0cm 
p ' = \frac{\hbar m_2}{2}, \hskip 1.0cm 
p'' = \frac{\hbar m_3}{2}
\label{52}
\end{equation} 
with $m_1,  m_2,  m_3$ are a new set of integers. 
Thus, up to a factor of half this gives  the  usual quantization rule for the
angular momentum. In other words, the star product automatically
incorporates the quantization condition even at the classical level.
Here too the exponent in (\ref{50}) can be written as the area of a phase space
triangle, in the cylindrical phase space mentioned earlier, with vertices 
${\bf r} = (\theta , p_{\theta}), {\bf r} ' = (\theta ' ,
 p')$ and ${\bf r} '' = (\theta '', p '') $, the only difference with 
the case of a particle on a line being that the momentum variables vary over
a discrete spectrum. 
That is 
\begin{equation}
- 2A({\bf r}, {\bf r} ', {\bf r} '') = p(\theta '' - \theta ') + p'(\theta - \theta '') +
p ''(\theta '- \theta ).
\label{53}
\end{equation}
Thus the star product in this case can be  written as
\begin{equation}
f \star g =  \frac{1}{(2\pi )^2}\sum_{m, n}\int_{-\pi}^{+\pi} d\theta '
 d \theta '' \int_{-\infty}^{+\infty} dp' dp'' \delta (p_{\theta} - p' +
\frac{\hbar m}{2} ) \delta(p_{\theta} - p'' - \frac{\hbar n}{2}  )
e^{[-\frac{4i}{\hbar}A({\bf r}, {\bf r} ', {\bf r} '')]}
f({\bf r}') g ({\bf r}'').
\label{54}
\end{equation}
Now we make a redefinition of the momentum variables as follows.
\begin{equation}
P_{\theta} =  \frac{2p_{\theta}}{\hbar}, \hskip 1.0cm P' = \frac{2p'}{\hbar},
\hskip 1.0 cm P '' = \frac{2p''}{\hbar}.
\label{55}
\end{equation}
Notice that this is not a canonical transformation. In terms of the new
 variables we can write $f \star g$ as
\begin{equation}
f \star g =  \frac{1}{(2\pi )^2}\sum_{m, n}\int_{-\pi}^{+\pi} d\theta '
 d \theta '' \int_{-\infty}^{+\infty} dP' dP'' \delta (P_{\theta} - P' +m)
\delta(P_{\theta} - P'' -n) e^{[-{2i}A(\tilde{\bf r}, 
\tilde{\bf r} ', \tilde{\bf r} '')]}
f(\tilde{\bf r}') g (\tilde{\bf r''}).
\label{56}
\end{equation}
Here $\tilde{\bf r}$ stands for $(\theta, P)$ now.
The form of the delta functions appearing in the above expression for 
star product enables one to replace the pair of Dirac delta functions
 with suitable Kronecker deltas and 
simultaneously changing integrations over $P', P''$ with summations to 
yield
\begin{equation}
f \star g =  \frac{1}{(2\pi )^2}\sum_{m, n, P', P''} \int_{-\pi}^{+\pi} 
d\theta 'd \theta ''  \delta_{P', P_{\theta}+m} \delta_{P'', P_{\theta}-n}
e^{[-2iA(\tilde{\bf r}, \tilde{\bf r} ', \tilde{\bf r} '')]}
f(\tilde{\bf r}') g (\tilde{\bf r''})
\label{57}
\end{equation}
Now on account of the restrictions imposed by the Kronecker deltas,
the summation over $P', P''$ implies the $m, n$ summations also.
Hence we obtain
\begin{equation}
(f \star g)(\tilde{\bf r}) =  \frac{1}{(2\pi )^2}\sum_{P', P''}\int_{-\pi}^{+\pi} d\theta '
d \theta '' e^{[-2iA(\tilde{\bf r}, \tilde{\bf r} ', \tilde{\bf r} '')]}
f(\tilde{\bf r}') g (\tilde{\bf r''})
\label{58}
\end{equation}
This expression is same  as (\ref{1-11}) which was obtained for the star product for a 1-d particle  except that  in place of the continuous variables $p', p''$,
 here we have  the discrete variables $P', P''$(defined in (\ref{55})
with a built-in factor of ($2$/$\hbar$)) and hence the corresponding
integrations are replaced by summations. Although the associativity is a built in property in the definition of star product, we
nevertheless carry out the verification here through the ``triangle" approach to indicate the similarities and differences of the case, where the particle is moving on the real line. 
The first step towards this is to express the product
$(f \star g) \star h$ in terms of the area of a phase space parallelogram.
To achieve this end, exploiting (\ref{58}) we write,
$$((f \star g) \star h)(\tilde{\bf r}) = \frac{1}{(2\pi )^2}\sum_{P', P''}\int_{-\pi}^{+\pi} d\theta '
d \theta '' e^{[-{2i}A(\tilde{\bf r}, \tilde{\bf r} ', \tilde{\bf r} '')]}
(f\star g )(\tilde{\bf r} ') h(\tilde{\bf r} '') 
$$
\begin{equation}
= \frac{1}{(2\pi )^4}\sum_{P', P'', P''', P''''}\int_{-\pi}^{+\pi} d\theta '
d \theta ''d \theta '''d \theta '''' e^{-{2i}[A(\tilde{\bf r},
\tilde{\bf r} ',
\tilde{\bf r} '')  + A(\tilde{\bf r'}, \tilde{\bf r} ''',
\tilde{\bf r} '''')]} f(\tilde{\bf r} ''') g(\tilde{\bf r} '''') h (\tilde{\bf r} '').
\label{59}
\end{equation} 
The exponent appearing in the above equation can be written as 
\begin{equation}
2A(\tilde{\bf r}, \tilde{\bf r} ',
\tilde{\bf r} '')  + 2A(\tilde{\bf r'}, \tilde{\bf r} ''',
\tilde{\bf r} '''') = \tilde{\bf r} ' \times [-\tilde{\bf r} + \tilde{\bf r} '' + \tilde{\bf r} ''' - 
\tilde{\bf r} '''' ] + \tilde{\bf r} '' \times\tilde{\bf r} +
 \tilde{\bf r} ''' \times  \tilde{\bf r} ''''.
\label{60}
\end{equation} 
Using (\ref{60}) and the relation
$$
\sum_{P'}\int_{-\pi}^{+\pi} d\theta ' e^{i\tilde{\bf r} ' \times [-\tilde{\bf r} + \tilde{\bf r} '' + \tilde{\bf r} ''' - \tilde{\bf r}
'''' ]} 
 = \sum_{P'}\int_{-\pi}^{+\pi} d\theta ' e^{i[\theta' (-P_\theta + P'' + P'''
 - P'''') - P'(-\theta + \theta '' + \theta ''' - \theta '''')]} $$
\begin{eqnarray}
& = & (2\pi)^2 \delta_{(-P_\theta + P'' + P'''  - P''''), 0}
 \delta_p(-\theta + \theta '' + \theta ''' - \theta '''')
\label{61}
\end{eqnarray}
we can write (\ref{59}) as
$$
((f \star g) \star h)(\tilde{\bf r}) = \frac{1}{(2\pi )^2} \sum_{ P'', P''', P''''}
\int_{-\pi}^{+\pi} d \theta ''d \theta '''d \theta ''''  \delta_{(P_{\theta} +
 P'' + P'''  - P''''), 0}
 \delta_p(-\theta + \theta '' + \theta ''' - \theta '''') $$
\begin{equation}
\times e^{-{i} 
[\tilde{\bf r} '' \times\tilde{\bf r} + \tilde{\bf r} ''' \times  \tilde{\bf r}
 '''']} f(\tilde{\bf r} ''') g(\tilde{\bf r} '''') h (\tilde{\bf r} '').
\label{62}
\end{equation} 
Comparing with (\ref{1-13}), we notice here the occurrence  of a product of a
Kroenecker delta and a periodic delta function instead of a product of two 
Dirac delta functions. 
The Kronecker delta present in the above product ensures that the $P$ sector
of the summation variables  indeed form the vertices of a parallelogram
though the same cannot be said of the $\theta$ part because of the periodic
delta function, as it cannot strictly enforce the requirement $(-\theta + \theta '' + \theta ''' - \theta '''') =0$. Nevertheless, the formal appearance of the 
above expression is enough to prove the associative of the star product.
Similar calculation for the product $f \star( g \star h)$ indeed gives an 
identical
result thereby proving associativity.  

\subsection{Particle on the Surface of a Sphere $S^2$}
We now consider the case of a particle constrained to the surface of a sphere.
The analysis in this case proceeds exactly as that of the particle on a circle
discussed before. 
The Lagrangian of the system and the corresponding constraints are given by
(\ref{32}),(\ref{33}) and (\ref{35}) with $i = 1, 2,3$. Similarly the 
Dirac bracket and its
kernel are obtainable from (\ref{37}) and (\ref{38}) respectively. 
In the present case,the $\{J^{\mu \nu}\}$ 
matrix involved in the kernel (\ref{38})  acts on the  6-dimensional phase 
space spanned by $(x_i, p_i)$ and is given by 
\begin{equation}
\{J^{\mu \nu}\} = \left( \begin{array}{cccccc}
0 & 0 & 0 &  1-x_1^2 & -x_1 x_2 & -x_1 x_3 \\
0 & 0 & 0 &  -x_1x_2 & 1- x_2^2 & -x_2x_3 \\
0 & 0 & 0 &  -x_1x_3 & -x_2x_3 & 1- x_3^2 \\
-(1-x_1^2) & x_1x_2 & x_1x_3 & 0 & x_2p_1 -x_1p_2 & x_3p_1 - x_1p_3\\
x_1x_2 & -(1-x_2^2) & x_2x_3 & x_1p_2 -x_2p_1 &  0 & x_3p_2 -x_2p_3 \\

x_1x_3 & x_2x_3 & -(1-x_3^2) & x_1p_3 - x_3p_1 & x_2p_3 -x_3p_2 & 0 
\end{array} \right).
\label{63}
\end{equation}
 As was done in the previous examples,
 we make a  transformation from the $(x_i, p_i)$ coordinate system
to a new system of independent canonical variables  $({\bar x},  \phi, \theta, {\bar p},  p_{\phi}, p_{\theta})$, 
\begin{eqnarray}
 \bar{x} = \frac{1}{\sqrt{2} }( x_i^2 - 1), &  \bar{p} = \frac{1}{\sqrt{2} }( x_ip_i) \\
{\phi} = tan^{-1} \frac{x_2}{x_1},  &  p_{\phi}  = x_1p_2 - x_2p_1 \\
\theta = cos^{-1} x_3, & p_{\theta} = \frac{x_3(x_i p_i) -
p_3(x_i^2)}{\sqrt{x_1^2 + x_2^2}}.
\label{64}
\end{eqnarray}
The canonical pair $({\bar x}, {\bar p})$ are proportional to the constraints and the    unconstrained variables are $(\phi, \theta, p_{\phi}, p_{\theta})$. 
The variables $\phi$ and $\theta$ are nothing but the azimuth and polar angles
respectively and hence have the compact ranges $0 \leq \phi < 2\pi$ and
$0 \leq \theta \leq \pi $. 
We introduce the notation $\phi = \theta_1,  \theta = \theta_2$ and 
proceed just as in the case of a particle on a circle, in the bracket kernel, 
we eliminate the 
partial differential operators with respect to the Cartesian coordinates
 in favor of those with respect to the new set of variables $\bar{x}, \bar{p}$
 and $ \theta_a,
p_{{\theta}_a}$ with $a= 1,2$.
 The corresponding bracket kernel reduces to the following
simple form
\begin{equation}
\Lambda = 
\sum_{a = 1, 2} \left(\lp_{{\theta}_a} \rp_{p_{{\theta}_a}} - 
\lp_{p_{{\theta}_a}} \rp_{{\theta}_a} \right) - \sqrt{2}{\bar x} \left(\lp_{\bar x}
\rp_{\bar p} - \lp_{\bar p} \rp_{\bar  x} \right)
\label{65}
\end{equation}
Upon implementing the constraint $1 -x_i^2 \approx 0$ strongly,  one finds that the variables $ {\bar x}$ and ${\bar p}$ disappear
from the bracket kernel(\ref{65}) and it 
reduces to the canonical form
\begin{equation}
\Lambda =
\sum_{a = 1, 2} \left(\lp_{{\theta}_a} \rp_{p_{{\theta}_a}} -
\lp_{p_{{\theta}_a}} \rp_{{\theta}_a} \right).
\label{66}
\end{equation}
which involves only the physical variables. This is again in conformity with
the Maskawa-Nakajima theorem.
Now it is obvious that  this kernel, when exponentiated, produces a
perfectly well behaved associative star product. 
Here we would like to emphasize once again, that algebraic verification of associativity can be carried out in each order 
of $\hbar$ just as for the particle moving on the real line, as the algebraic structure of the bracket kernel in either case takes
a similar canonical form and this is regardless of the nature of these coordinates. Also, the fact that the coordinate chart
$(\theta , \phi )$ does not cover the entire $S^2$  is of no consequence as we can define new polar coordinates with 
respect to an $SO(3)$ rotated axes or stereographic variables as was done for $S^1$ to define another chart. Two or more compatible charts 
can be used to provide an atlas for the whole of $S^2$. And the bracket kernel can be shown to take a similar canonical form
in all the charts. We have demonstrated this in the case of $S^1$. It can be easily demonstrated for $S^2$
also.

Finally let us make the following observations. 
At the classical level,
the spherical polar coordinate variables $\theta$ and $\phi$ 
 having the compact
ranges $ 0 \leq \theta \leq \pi $ and $0 \leq \phi < 2 \pi$ respectively,
coordinatize only a neighborhood of the 
 manifold $S^2$ whereas their
 conjugate momenta $p_{\theta}, p_{\phi}$  varying from 
$-\infty$ to $+\infty$ constitute a plane ${\cal R}^2$. Thus  the  phase space
of the system  given by $(\theta , \phi , p_{\theta} , p_{\phi})$ cover only a neighborhood of the whole phase space. 
Besides, the basis  for the
expansion (analogous to (\ref{1}) and (\ref{47}))  of functions in $S^2 $ are
the spherical harmonics $Y_{lm} (\theta, \phi)$ while in 
${\cal R}^2$ the usual exponential functions provide the basis. Though the 
$\phi$ dependence of $Y_{lm} (\theta, \phi)$ is via the usual exponential functions, its $\theta$ dependence is through
the associated Legendre polynomials $P_{lm} (\cos\theta)$ which are not 
eigenfunctions of 
$\partial_{\theta}$. This suggests that carrying out a naive Fourier/harmonic analysis may not yield any quantization condition
for the angular momentum as could be done for $S^1$. The coordinate independent Fourier transform of \cite{batalin}
may be useful for this purpose. This nontrivial point requires further investigation.

Unlike the previous
examples, the Groenewold method of reduction of the star product to an integral involving
the exponentiated area of a phase space triangle(i.e. analogous to (\ref{13})
and (\ref{58}))  is not viable for particle
on $S^2 $. The source of this difficulty can be traced to the nontrivial geometry and/or topology 
of the phase space.
But as has been shown in \cite{batalin}, the construction of the star product has built-in associativity and 
triangle interpretation is also available, where one gets a geodesic triangle which reduces to a rectilinear triangle
(as in the non-degenerate case) in terms of the coordinates on the constraint surface. 
In the phase space approach of Weyl-Groenewold \cite{groenewold, weyl}, demonstration of associativity is 
done by using the MN variables as discussed here.
\section{Conclusion}

In this paper we  studied  various aspects of the construction of star
product for certain physical systems. The construction and properties of 
star product were reviewed in the case of a particle moving on a real line. 
 We have shown how an associative star 
product can be constructed for systems involving nonlinear constraints. 
As is well known, the Dirac brackets in this case are variable dependent 
and a naive exponentiation of the bracket kernel does not yield a manifestly associative 
star product. By exploiting a theorem due to Maskawa and Nakajima(MN) \cite{mn},
we were able to extract the physical unconstrained set of variables. 
The bracket kernel, computed with respect to these variables, simplified to the
standard canonical symplectic structure, once the constraints were imposed.
Associativity of the star product is then manifestly evident.  

Our method was illustrated with the examples of Landau problem, particle on a circle and
a particle restricted to be on the surface of a sphere.  The Kontsevich series
 expansion of the star product was shown to reduce to  an exponential series
when expressed in terms of the unconstrained variables of the systems. 
Similar to the case of a 1-dimensional free particle, we showed that
the associativity of the star product for a particle on the circle can
be given the triangle (in the cylindrical phase space) interpretation 
with the difference that the angular momentum taking integral values up to a 
factor of ($\hbar$/$2$).
We thus obtained the usual quantization condition 
of the angular momentum (up to a factor of half) in the case of a particle 
moving on the circle.

The quantization condition of the angular momentum,  however, could not be reproduced for a particle
moving on the surface of a sphere.
Also,  MN variables
can  be defined only locally for systems with  topologically more complicated
 phase spaces (particle constrained on a sphere for example).  
Also, in the case of ${\cal S}^2$  (as for most degenerate systems), the star product cannot be given the
triangle interpretation in the Weyl-Groenewold approach. Nevertheless,
it must be emphasized that the lack of  such an interpretation or
the local nature of the MN variables are of no consequence in demonstrating
the associativity of the star product using MN variables. The reason for this is that 
 we employ a purely {\it algebraic} method for demonstrating associativity
rather than a geometric one. This was illustrated clearly in the case of the
examples considered in this paper, viz., Landau problem, particles moving on a
circle and on a sphere.  

It may be  mentioned that, in the coordinate free formulation of
 Batalin and Tyutin \cite{batalin}, the star product is defined in such a
way as to satisfy associativity property automatically. Thus it is always
possible to give a 
triangle interpretation to the star product irrespective of the geometry the
base manifold. For degenerate systems with second class constraints, the star product
can be written in terms of the exponential of the area of a geodesic triangle which
reduces to a rectilinear triangle on the constraint  surface. Thus, a geometric 
interpretation of  the associativity of star product is always available in the 
coordinate independent approach  for both degenerate and non-degenerate systems.

Finally, we feel that the coordinate independent Fourier transform of \cite{batalin} would be helpful in 
resolving   
the difficulties mentioned in section 4.3 for giving an area interpretation for the associativity of 
star product(phase space approach) and obtaining the angular momentum quantization 
condition  for a particle on the surface of a sphere.  
This problem
is under further investigation and will be reported elsewhere.


\begin{thebibliography}{99}
\bibitem{connes}A. Connes, {\it Noncommutative Geometry}, Academic Press, San Diego(1994). 
\bibitem{bayen}F. Bayen, M. Flato, C. Fronsdal, A. Lecherenowicz and 
D. Sternheimer, {\it Ann. Phys.} {\bf 111} (1978) 61.
\bibitem{zachos} C. Zachos,  archive report hep-th/0008010.
\bibitem{groenewold} H. Groenewold, {\it Physica }{\bf 12} (1946) 405.
\bibitem{moyal} J. Moyal, {\it Proc. Camb.  Phil. Soc}, {\bf 45} (1949) 99.
\bibitem{berezin} F. A. Berezin, {\it Comm. Math. Phys.} {\bf 40} (1975) 153.
\bibitem{batalin} I. A. Batalin and I. V. Tyutin, {\it Nucl. Phys. } {\bf B345} (1990) 645.  
\bibitem{witten}N. Seiberg and E. Witten, {\it JHEP} {\bf 9909} (1999) 032.
\bibitem{szabo}R.J. Szabo, archive report hep-th/0109162.
\bibitem{dn}M.R. Douglas and N.A. Nekrasov, {\it Rev.Mod.Phys.} {\bf 73} (2002) 977.
\bibitem{bcg} R. Banerjee, B. Chakraborty and S. Ghosh, {\it Phys. Lett.} {\bf B537} (2002) 340. 
\bibitem{kontsevich} M. Kontsevich, archive report q-alg/9709040.
\bibitem{mn} T. Maskawa and H. Nakajima, {\it Prog.Theor.Phys.} {\bf 56} (1976) 1295.
\bibitem{weyl} H. Weyl, {\it Z.  Phys. }{\bf 46} (1927) 1. 
\bibitem{zachos2} C. Zachos,  {\it Int.J.Mod.Phys.}{\bf A17} (2002) 297. 
\bibitem{weinberg}S. Weinberg, {\it The Quantum Theory of Fields. Vol. 1}
Cambridge University Press, New York(1996).
\bibitem{dirac}P.A.M. Dirac, {\it Lectures on Quantum Mechanics}, Belfer
Graduate School of Sciences, Yeshiva University, New York(1964). 
\end{thebibliography}
\end{document}